\documentclass[runningheads]{llncs}
\usepackage[T1]{fontenc}
\usepackage{graphicx}
\usepackage{amssymb}
\usepackage{amsmath}
\usepackage{multirow}
\usepackage[misc]{ifsym}
\usepackage{hyperref}
\begin{document}
%
\title{Flow-based Visual Quality Enhancer for Super-resolution Magnetic Resonance Spectroscopic Imaging}
\titlerunning{Flow-based Visual Quality Enhancer for Super-resolution MRSI}
%
\author{Siyuan Dong\inst{1}\textsuperscript{(\Letter)} \and Gilbert Hangel\inst{2} \and Eric Z. Chen\inst{3} \and Shanhui Sun\inst{3} \and Wolfgang Bogner\inst{2} \and Georg Widhalm\inst{4} \and Chenyu You\inst{1} \and John A. Onofrey\inst{5} \and Robin de Graaf\inst{5} \and James S. Duncan\inst{1,5}}

%
\authorrunning{S. Dong et al.}
%
%
\institute{Electrical Engineering, Yale University, New Haven, CT, USA
\email{s.dong@yale.edu} \and Biomedical Imaging and Image-guided Therapy, Highfield
MR Center, Medical University of Vienna, Vienna, Austria \and
United Imaging Intelligence, Cambridge, MA, USA \and Neurosurgery, Medical University of Vienna, Vienna, Austria \and Radiology and Biomedical Imaging, Yale University, New Haven, CT, USA}
\maketitle              
\begin{abstract}
Magnetic Resonance Spectroscopic Imaging (MRSI) is an essential tool for quantifying metabolites in the body, but the low spatial resolution limits its clinical applications. Deep learning-based super-resolution methods provided promising results for improving the spatial resolution of MRSI, but the super-resolved images are often blurry compared to the experimentally-acquired high-resolution images. Attempts have been made with the generative adversarial networks to improve the image visual quality. In this work, we consider another type of generative model, the flow-based model, of which the training is more stable and interpretable compared to the adversarial networks. Specifically, we propose a flow-based enhancer network to improve the visual quality of super-resolution MRSI. Different from previous flow-based models, our enhancer network incorporates anatomical information from additional image modalities (MRI) and uses a learnable base distribution. In addition, we impose a guide loss and a data-consistency loss to encourage the network to generate images with high visual quality while maintaining high fidelity. Experiments on a \textsuperscript{1}H-MRSI dataset acquired from 25 high-grade glioma patients indicate that our enhancer network outperforms the adversarial networks and the baseline flow-based methods. Our method also allows visual quality adjustment and uncertainty estimation. Our code is available at \url{https://github.com/dsy199610/Flow-Enhancer-SR-MRSI}. 

\end{abstract}
\section{Introduction}
Magnetic Resonance Spectroscopic Imaging (MRSI) is a technique for measuring metabolite concentrations within the body \cite{de2019vivo}. Because the metabolic level is a crucial indicator of cell activities, MRSI is becoming a valuable tool for studying different diseases such as brain tumors \cite{hangel2020high} and cancers \cite{coman2020extracellular}. However, due to the low concentrations of metabolites, current applications of MRSI are limited to low spatial resolutions. Hence, developing a post-processing algorithm for generating higher resolution MRSI will greatly benefit its clinical applications. 

Recent advances in deep learning have provided promising results for super-resolution (SR) MRSI \cite{cengiz2017super,iqbal2019super}. These works trained neural networks to map low-resolution MRSI metabolic maps to higher-resolution ones with a pixelwise mean-squared error (MSE) loss function. However, SR is a one-to-many problem, and training with MSE learns a pixelwise average of all possible solutions \cite{menon2020pulse,li2021best}. This can result in blurry SR images with suboptimal visual quality and a lack of high-frequency details \cite{dong2022invertible}. To approach this issue, a few recent works proposed to add adversarial loss to improve the visual quality \cite{dong2021high,li2022deep,dong2022multiscale}, but it is well-known that the generative adversarial networks suffer from training instability and mode collapse \cite{lugmayr2020srflow,ardizzone2018analyzing}. The normalizing flow (NF) model is a relatively new class of generative models that learns the target distribution via the maximization of exact log-likelihood, making the training more interpretable and stable \cite{dinh2016density,kingma2018glow,lugmayr2020srflow,ardizzone2018analyzing}. NF learns an invertible mapping from the target image distribution to a simple base distribution during training, so the target images can be generated by sampling from the base distribution during inference. A few recent works applied NF on SR of natural images \cite{lugmayr2020srflow,liang2021hierarchical} and reconstructions of medical images \cite{denker2021conditional,kelkar2021compressible}, and these flow-based methods can generate images with high visual quality.

In this work, we propose a flow-based enhancer network to recover high-frequency details and improve the visual quality of the blurry SR MRSI images given by the SR networks. Borrowing the idea of sharpness enhancement \cite{dong2022invertible}, we regard the visual quality enhancement as a subsequent step of the SR network. Hence, the enhancer network only needs to focus on improving the visual quality, not the entire SR process. To boost the performance, we make several modifications to the existing flow-based SR method \cite{lugmayr2020srflow}, including incorporating MRI anatomical information and a learnable base distribution. We also enforce a data-consistency (DC) loss to encourage the enhanced images not to alter the original measurements from the scanner, which is very important for medical images. Experimental results show that our enhancer network successfully improves the visual quality of SR metabolic maps, outperforming the adversarial networks and the baseline flow-based methods. Our method also allows visual quality adjustment and uncertainty estimation within the same network.

\section{Methods}

\begin{figure}[t]
\centering
\includegraphics[width=0.99\textwidth]{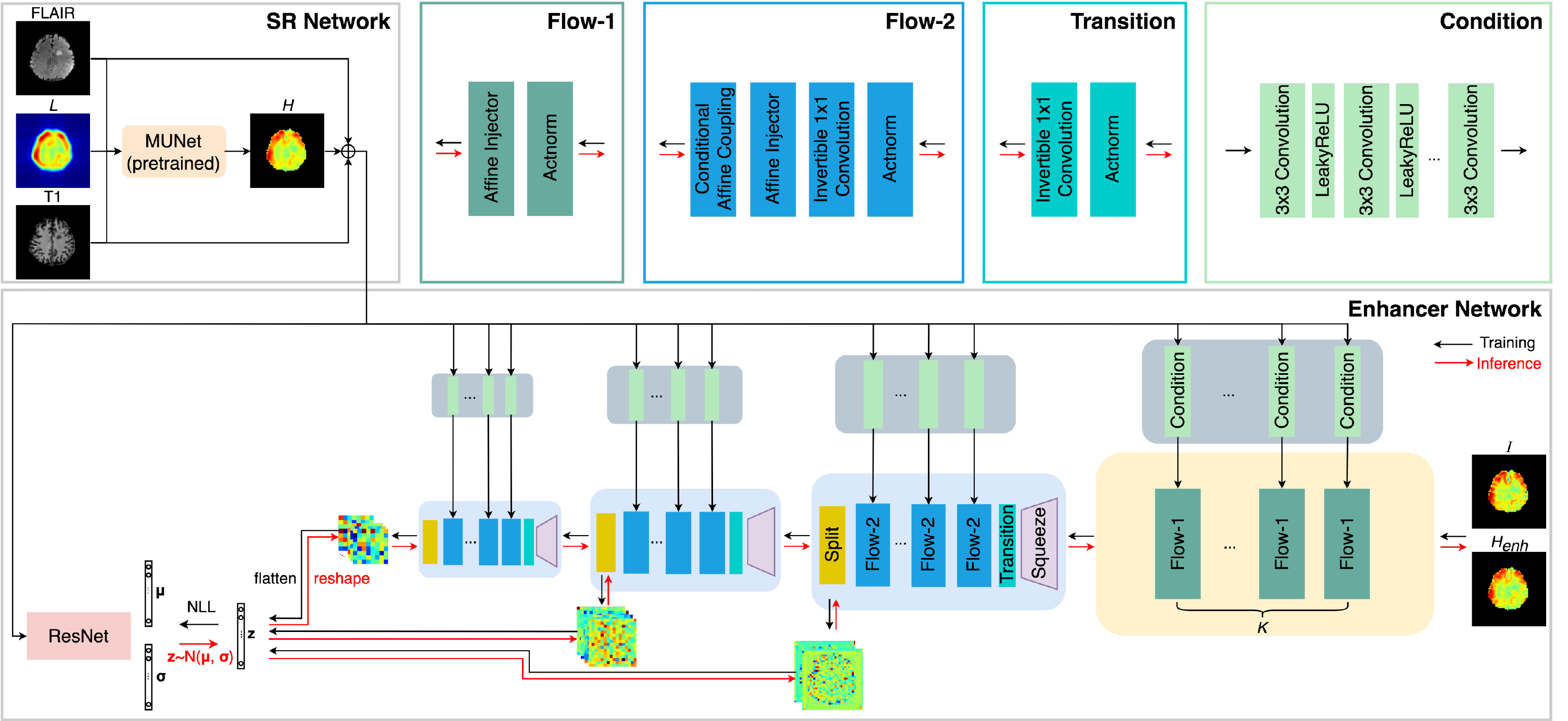}
\caption{Proposed flow-based enhancer network for improving the visual quality of SR MRSI. We first use a pretrained Multi-encoder UNet (MUNet) \cite{dolz2018dense,dong2021high,dong2022multiscale} as the SR Network to obtain a super-resolved image $H$ from the low-resolution image $L$. The enhancer network takes \{$H$, T1, FLAIR\} as the condition, which is processed with the Condition blocks before feeding into the flow layers. The flow layers Flow-2 and Transition follow the design in SRFlow \cite{lugmayr2020srflow}. We add a special type of flow layer, Flow-1, before any squeeze operation to process the image in its original dimension \cite{padmanabha2021solving}. During training, the enhancer network transforms the ground truth images $I$ into the Gaussian vectors $\mathbf{z}$ using the NLL loss. During inference, $\mathbf{z}$ is drawn from the Gaussian distribution and inversely passed through the network to obtain $H_{enh}$. The mean $\boldsymbol{\mu}$ and standard deviation $\boldsymbol{\sigma}$ of the Gaussian distribution is learned from the condition \{$H$, T1, FLAIR\} using a ResNet \cite{he2016deep}.} 
\label{fig1}
\end{figure}

\subsection{Problem Formulation}
Given a low-resolution metabolic map $L \in \mathbb{R}^{n\times n}$, the SR network $S$ learns a mapping $H = S(L)$ such that the super-resolved metabolic map $H \in \mathbb{R}^{N\times N}$ is close to the ground truth high-resolution map $I \in \mathbb{R}^{N\times N}$. However, the SR network $S$ trained with pixelwise loss \cite{cengiz2017super,iqbal2019super} and structural loss \cite{dong2021high} could result in blurry $H$. We develop an enhancer network $E$ to improve the visual quality of $H$, i.e. $H_{enh} = E(H)$, such that $H_{enh}$ has visual quality comparable to $I$. 

\subsection{Preliminary: Conditional Normalizing Flow}
NF is a family of generative models that constructs a complex distribution from a simple distribution using a flow of invertible transformations \cite{rezende2015variational}. The key idea is to learn a bijective mapping between the target space and a latent space \cite{dinh2016density}. Given that $\mathbf{x}$ is a sample from the target space with distribution $p_{\mathbf{x}}(\mathbf{x})$, flow-based models usually use an invertible neural network $f$ to transform $\mathbf{x}$ into a latent variable $\mathbf{z}$ with a simple base distribution $p_{\mathbf{z}}(\mathbf{z})$, e.g. Gaussian. Once the transformation is learned, the network $f$ can generate samples from the target distribution $\mathbf{x}=f^{-1}(\mathbf{z})$ by sampling $\mathbf{z} \sim p_{\mathbf{z}}(\mathbf{z})$. This idea can also be used to learn a conditional distribution $p_{\mathbf{x}|\mathbf{y}}(\mathbf{x}|\mathbf{y})$ over two random variables $\mathbf{x}$ and $\mathbf{y}$, also known as the conditional NF \cite{winkler2019learning,lugmayr2020srflow}. According to the change of variable formula, the target distribution $p_{\mathbf{x|y}}(\mathbf{x|y})$ can be expressed as
\begin{equation}
    p_{\mathbf{x|y}}(\mathbf{x|y}) = p_{\mathbf{z}}(f(\mathbf{x;y}))\left|\text{det}\left(\frac{\partial f(\mathbf{x;y})}{\partial \mathbf{x}}\right)\right|
\end{equation}
where $\frac{\partial f(\mathbf{x;y})}{\partial \mathbf{x}}$ is the Jacobian matrix. This expression allows training the network $f$ with the negative log-likelihood (NLL) loss for training samples $(\mathbf{x}, \mathbf{y})$
\begin{equation}
    \mathcal{L}_{NLL} = -\text{log} \; p_{\mathbf{x|y}}(\mathbf{x|y}) = -\text{log} \; p_{\mathbf{z}}(f(\mathbf{x;y})) - \text{log} \; \left|\text{det}\left(\frac{\partial f(\mathbf{x;y})}{\partial \mathbf{x}}\right)\right|.
\label{eqn_NLL}
\end{equation}
In this work, we train our enhancer network $E$ with NLL loss to learn the ground truth space conditioned on the corresponding blurry SR image, i.e. to learn $p_{I|H}$.

\subsection{Network Architecture}
Our enhancer network (Fig. \ref{fig1}) adopts a multi-scale architecture \cite{dinh2016density,kingma2018glow,lugmayr2020srflow}. Each scale consists of a series of fully invertible conditional flow layers with tractable Jacobian, so the NLL loss in Equation \ref{eqn_NLL} can be computed. We use the same flow layers as in SRFlow \cite{lugmayr2020srflow}, specifically, conditional affine coupling, affine injector, invertible $1 \times 1$ convolution and activation normalization (Actnorm) \cite{kingma2018glow}. Our enhancer network is different from the SRFlow network in mainly three aspects: (1) SRFlow is a SR network that super-resolves the low-resolution images, whereas our method enhances the super-resolved images given by any SR network that are blurry; (2) SRFlow processes a single image modality, but our network incorporates information from other modalities (T1 and FLAIR MRI); (3) SRFlow uses a fixed base distribution, standard Gaussian $p_{\mathbf{z}}=\mathcal{N}(0, I)$, but our network adopts a learnable base distribution. 

\textbf{MRI Anatomical Information} Previous works suggested that the multi-parametric MRI contains useful prior information for SR MRSI \cite{dong2021high,dong2022multiscale}. Therefore, we provide T1 and FLAIR MRI as additional conditions to the enhancer network. Specifically, T1 and FLAIR are re-sampled and concatenated with the super-resolved image $H$ (we denote this as $\{H$, T1, FLAIR$\}$), which are then passed into the Condition blocks of the enhancer network (see Fig. \ref{fig1}). 

\textbf{Learnable Base Distribution} Current flow-based SR models \cite{lugmayr2020srflow,liang2021hierarchical} assume a $0$-mean and unit-norm multivariate Gaussian for the base distribution, whereas such a predetermined base distribution might limit the learning capability of the model. We modify it as a multivariate Gaussian with learnable mean and standard deviation, i.e. $p_{\mathbf{z}}=\mathcal{N}(\boldsymbol{\mu}(c), \boldsymbol{\sigma}(c))$, where the mean and standard deviation vectors are learned from the condition $c$=$\{H$, T1, FLAIR$\}$ using a ResNet \cite{he2016deep}. This ``conditional base'' was included in the original conditional NF work \cite{winkler2019learning} but was neglected in later applications. 

\subsection{Loss Function}
We apply the NLL loss to learn the conditional distribution of ground truth images $I$ given the SR images $H$. Therefore, the NLL loss in Equation \ref{eqn_NLL} becomes
\begin{equation}
\begin{aligned}
    \mathcal{L}_{NLL}(I) &= -\text{log} \; p_{\mathbf{z}}(E(I;H)) - \text{log} \; \left|\text{det}\left(\frac{\partial E(I;H)}{\partial I}\right)\right| \\
    &= -\frac{1}{2}(\left\lVert\frac{\mathbf{z} - \boldsymbol{\mu}(c)}{\boldsymbol{\sigma}(c)}\right\rVert_{2}^{2} + \left\lVert\text{log}(2\pi \boldsymbol{\sigma}(c)^2)\right\rVert_{1}) - \text{log} \; \left|\text{det}\left(\frac{\partial E(I;H)}{\partial I}\right)\right|.
\label{loss_NLL}
\end{aligned}
\end{equation}
We define $\mathbf{z}=E(I;H)$ as the training direction and $H_{enh}=E^{-1}(\mathbf{z};H)$ as the inference direction. 
The second equation holds because $p_{\mathbf{z}}=\mathcal{N}(\boldsymbol{\mu}(c), \boldsymbol{\sigma}(c))$. The log-determinant term can be computed efficiently, because the flow layers are designed to have tractable Jacobian \cite{kingma2018glow}. 

We add a guide loss in the inference direction to guide the network to learn an enhanced image space centered around the fidelity-oriented SR image \cite{liang2021hierarchical}
\begin{equation}
    \mathcal{L}_{guide}(I, H_{enh}^{\tau=0}) = (1-\alpha) \mathcal{L}_{pixel}(I, H_{enh}^{\tau=0}) + \alpha \mathcal{L}_{structural}(I, H_{enh}^{\tau=0})
\end{equation}
where the temperature $\tau$ is a scale parameter that controls the variance of the random sample: $H_{enh}^{\tau = \tau_{0}} = E^{-1}(\mathbf{z};H)$ with $\mathbf{z} \sim \mathcal{N}(\boldsymbol{\mu}(c), \tau_{0} \boldsymbol{\sigma}(c))$. The pixel loss $L_{pixel}$ defines a pixelwise difference between two images using L1-norm, and the structural loss $L_{structural}$ maximizes the Multiscale Structural Similarity (MS-SSIM) \cite{wang2003multiscale} between two images \cite{dong2021high,dong2022multiscale}. 

Furthermore, we use a DC loss \cite{chen2020data} to encourage that the enhanced image follows the k-space measurement from the scanner
\begin{equation}
    \mathcal{L}_{DC}(L, H_{enh}^{\tau=\tilde{\tau}}) = \left\lVert \mathcal{F}(L) - \mathcal{F}_{u}(H_{enh}^{\tau=\tilde{\tau}})\right\rVert_{1}
\end{equation}
where $\mathcal{F}$ is the Fourier transform operator, and $\mathcal{F}_{u}$ denotes down-sampling after Fourier transform to match the dimension of the low-resolution measurement. $\tilde{\tau}$ is uniformly generated $\tilde{\tau} \sim \mathcal{U}(0, 1)$ during training to encourage DC at all temperature levels. DC loss instructs the enhancer network not to modify the scanner measurement, which is very important for reliable clinical diagnosis. 

Overall, the enhancer network is trained with:
\begin{equation}
    \mathcal{L} = \mathcal{L}_{NLL}(I) + \lambda_{1}\mathcal{L}_{guide}(I, H_{enh}^{\tau=0}) + \lambda_{2}\mathcal{L}_{DC}(L, H_{enh}^{\tau=\tilde{\tau}})
\end{equation}

\section{Experiments}
\subsection{Data Acquisition and Preprocessing}
We acquired 3D \textsuperscript{1}H-MRSI, T1 and FLAIR from 25 high-grade glioma patients using a Siemens 7T whole-body-MR imager \cite{hangel2020high}. IRB approval and informed consent from all participants were obtained. MRI images were skull-stripped using FSL v5.0 \cite{smith2004advances}. The MRSI sequences were acquired using an acquisition delay of 1.3 ms, repetition time of 450 ms and scan duration of 15 min. The nominal resolution is 3.4$\times$3.4$\times$3.4 mm$^3$, and the matrix size is 64$\times$64$\times$39. Note that this is a very high resolution for MRSI because of the challenges in acquiring metabolite signals with acceptable SNR. The MRSI spectra were quantified using LCModel v6.3-1 \cite{provencher2014lcmodel} to obtain the 3D metabolic maps. The voxels with insufficient quality (SNR $<$ 2.5 or FWHM $>$ 0.15 ppm) or under strong distortion around the brain periphery were excluded. We focused on 7 metabolites that are major markers of onco-metabolism \cite{hangel2020high}, namely N-acetyl-aspartate (NAA), total creatine (tCr), total choline (tCho), glutamate (Glu), glutamine (Gln), inositol (Ins), and glycine (Gly). 

\subsection{Implementation Details}
From every 3D MRSI scan, we obtained 9-18 axial slices, and each includes 7 metabolites, summing to 2275 2D metabolic maps. These are regarded as the high-resolution ground truth $I \in \mathbb{R}^{64\times 64}$, which were truncated in k-space to obtain low-resolution images $L \in \mathbb{R}^{16\times 16}$. Of the 25 patients, we used 15 for training, 5 for validation and 5 for testing. This corresponds to 1246 metabolic maps for training, 483 for validation and 546 for testing. The training dataset was augmented using random rotation and shifting at every training iteration. 

The enhancer network operates in 4 scales, each contains $K=12$ flow steps. The loss weighting parameters are $\alpha=0.84$ \cite{zhao2016loss}, $\lambda_{1}=10$ and $\lambda_{2}=10$. The experiments were implemented in PyTorch v1.1.0 and performed on NVIDIA GTX 1080 and V100 GPUs. The networks were trained with the Adam optimizer \cite{kingma2014adam}, initial learning rate of $1\times10^{-4}$, batch size of 8 and 500 epochs.  

\subsection{Results}

\begin{table}[t]
\centering
\caption{Quantitative results. Results are presented in mean $\pm$ standard deviation. A lower LPIPS score means better results ($\downarrow$). The best scores are shown in bold. }\label{tab1}
\begin{tabular}{|c |@{\hskip 0.01in}c@{\hskip 0.01in} |@{\hskip 0.01in}c@{\hskip 0.01in} |@{\hskip 0.01in}c@{\hskip 0.01in} |@{\hskip 0.01in}c@{\hskip 0.01in} |}
\hline 
Type & Methods & PSNR($\uparrow$) & SSIM($\uparrow$) & LPIPS($\downarrow$) \\
\hline \hline 
Fidelity-oriented & MUNet\cite{dong2021high} & \textbf{29.7 $\pm$ 2.5} & \textbf{0.933 $\pm$ 0.028} & 0.0897 $\pm$ 0.0462 \\
\hline 
 \multirow{3}*{Visual-oriented} & MUNet-cWGAN\cite{dong2021high} & 28.3 $\pm$ 2.6 & 0.920 $\pm$ 0.028 & 0.0529 $\pm$ 0.0349 \\
 & SRFlow\cite{lugmayr2020srflow} & 27.8 $\pm$ 2.5 & 0.905 $\pm$ 0.048 & 0.0656 $\pm$ 0.0516 \\
 & Flow Enhancer(ours) & 29.0 $\pm$ 2.4 & 0.924 $\pm$ 0.029 & \textbf{0.0519 $\pm$ 0.0340} \\
\hline
\end{tabular}
\end{table}

\begin{figure}[t]
\centering
\includegraphics[width=0.95\textwidth]{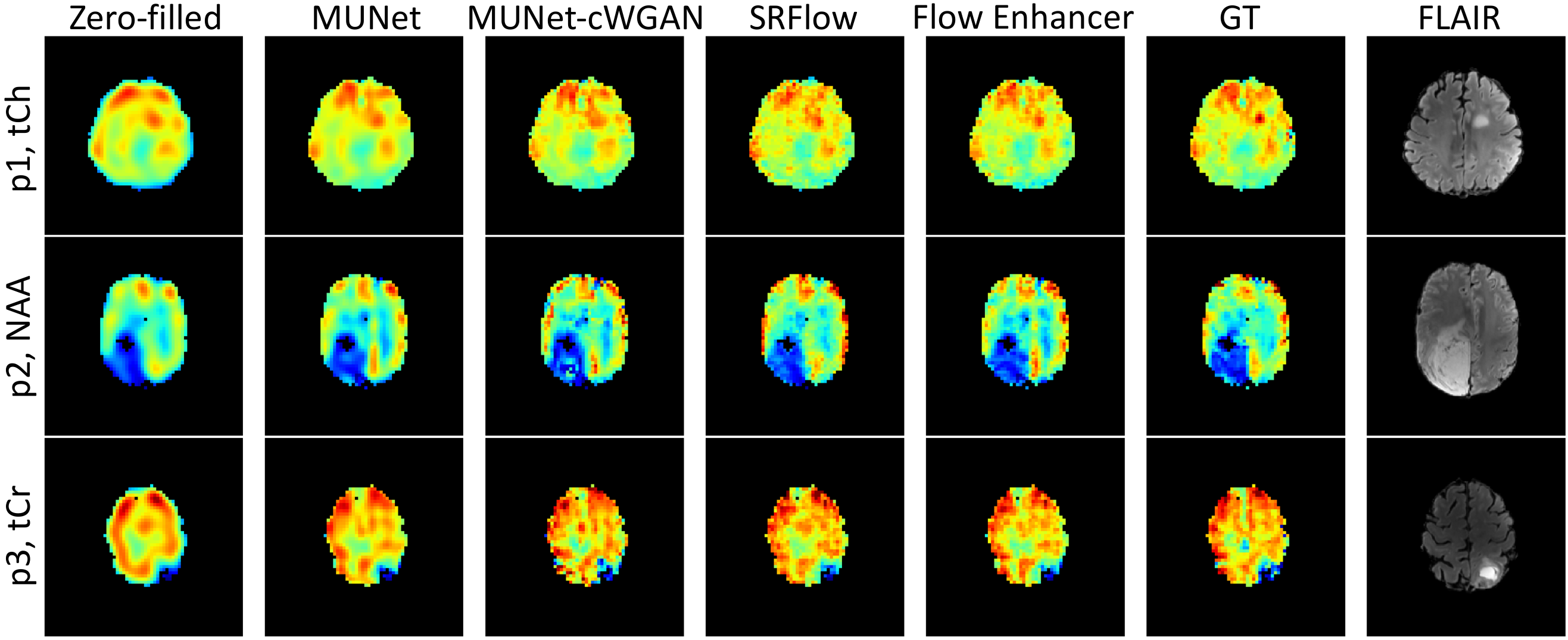}
\caption{Qualitative results. From left to right: k-space zero-filled images, MUNet, MUNet-cWGAN, SRFlow, our Flow Enhancer, ground truth (GT) and FLAIR images for anatomical reference. From top to bottom: tCh, NAA and tCr maps from three different patients p1, p2 and p3, respectively. Zoom in to inspect more details.} 
\label{fig2}
\end{figure}

We implemented the enhancer network to improve the visual quality of the SR images given by a Multi-encoder UNet (MUNet) \cite{dolz2018dense,dong2021high,dong2022multiscale}, which was trained with a pixelwise plus structural loss. The MUNet uses two encoders for processing anatomical information in T1 and FLIAR respectively. For comparison, we implemented two previous visual-oriented SR methods (these methods were applied to $L$): (1) MUNet trained with the adversarial loss using conditional Wasserstein generative adversarial networks (MUNet-cWGAN) \cite{dong2021high}, and (2) the baseline flow-based SR model that does not have our new design elements (e.g. MRI condition, $\mathcal{L}_{DC}$), denoted as SRFlow \cite{lugmayr2020srflow}. We set $\tau=0.8$ for the flow-based methods as recommended by previous works \cite{lugmayr2020srflow,liang2021hierarchical}. As for the evaluation metrics, peak signal-to-noise ratio (PSNR) and structural similarity index (SSIM) are the most commonly used metrics for evaluating image SR, but they are ineffective in measuring image visual quality \cite{dong2022invertible}. Previous literature indicates that these fidelity-oriented metrics are often degraded as the visual quality improves \cite{blau2018perception,lugmayr2020srflow,wang2020deep}. Here we report a visual-oriented metric, Learned Perceptual Image Patch Similarity (LPIPS), which measures the high-level similarity between two images using a pretrained deep network (AlexNet) and correlates well with human perceptual judgment \cite{zhang2018unreasonable,dong2022invertible,dong2022multiscale}. Table \ref{tab1} shows that although MUNet achieves high PSNR and SSIM scores, its LPIPS score is relatively poor compared to the visual-oriented methods. Compared to MUNet-cWGAN and SRFlow, our method (Flow Enhancer) achieves better visual quality (LPIPS) while maintaining higher fidelity (PSNR and SSIM). Fig. \ref{fig2} shows the corresponding qualitative comparisons. MUNet provides SR images that are blurry compared to the ground truth images. MUNet-cWGAN improves the visual quality but tends to generate more artifacts than our Flow Enhancer. In addition, as shown in the first row (p1, tCh), Flow Enhancer recovers better contrast at the tumor than SRFlow and MUNet-cWGAN. 

\begin{table}[t]
\centering
\caption{Ablation Studies. $\checkmark$ or $\times$ represents whether a certain design element is present or not. Paired t-test was performed between our method (last row) and the first four rows. Statistically significant differences ($p$-value $<0.05$) are shown with *. }\label{tab2}
\begin{tabular}{|c |c |c |c |c |c |c |}
\hline 
MRI prior & Cond. Base & $\mathcal{L}_{guide}$ & $\mathcal{L}_{DC}$ & PSNR($\uparrow$) & SSIM($\uparrow$) & LPIPS($\downarrow$) \\
\hline
$\times$ & $\checkmark$ & $\checkmark$ & $\checkmark$ & 28.9 $\pm$ 2.3* & 0.920 $\pm$ 0.034* & 0.0558 $\pm$ 0.0367* \\
\hline
$\checkmark$ & $\times$ & $\checkmark$ & $\checkmark$ & \textbf{29.0 $\pm$ 2.4}* & \textbf{0.924 $\pm$ 0.029} & 0.0526 $\pm$ 0.0337* \\
\hline
$\checkmark$ & $\checkmark$ & $\times$ & $\times$ & 28.3 $\pm$ 2.3* & 0.918 $\pm$ 0.029* & 0.0579 $\pm$ 0.0398* \\
\hline
$\checkmark$ & $\checkmark$ & $\checkmark$ & $\times$ & 28.8 $\pm$ 2.3* & 0.922 $\pm$ 0.029* & \textbf{0.0513 $\pm$ 0.0339}* \\
\hline
$\checkmark$ & $\checkmark$ & $\checkmark$ & $\checkmark$ & \textbf{29.0 $\pm$ 2.4} & \textbf{0.924 $\pm$ 0.029} & 0.0519 $\pm$ 0.0340 \\
\hline
\end{tabular}
\end{table}

\textbf{Ablation Studies} To justify our design, we performed ablation studies on 4 design elements: MRI anatomical information (MRI prior), conditional base distribution (Cond. Base), guide loss $\mathcal{L}_{guide}$ and DC loss $\mathcal{L}_{DC}$. Table \ref{tab2} indicates that removing the MRI prior (in this case, $c$=$\{H\}$) harms all three metrics. Removing $\mathcal{L}_{guide}$ degrades the performance by a large margin. Removing the Cond. Base does not harm PSNR/SSIM but gives slightly worse LPIPS. $\mathcal{L}_{DC}$ imposes an L1 loss on the low-resolution components in k-space, therefore adding $\mathcal{L}_{DC}$ gives better PSNR/SSIM but slightly sacrifices the LPIPS (visual quality). 

\begin{figure}
\centering
\includegraphics[width=0.91\textwidth]{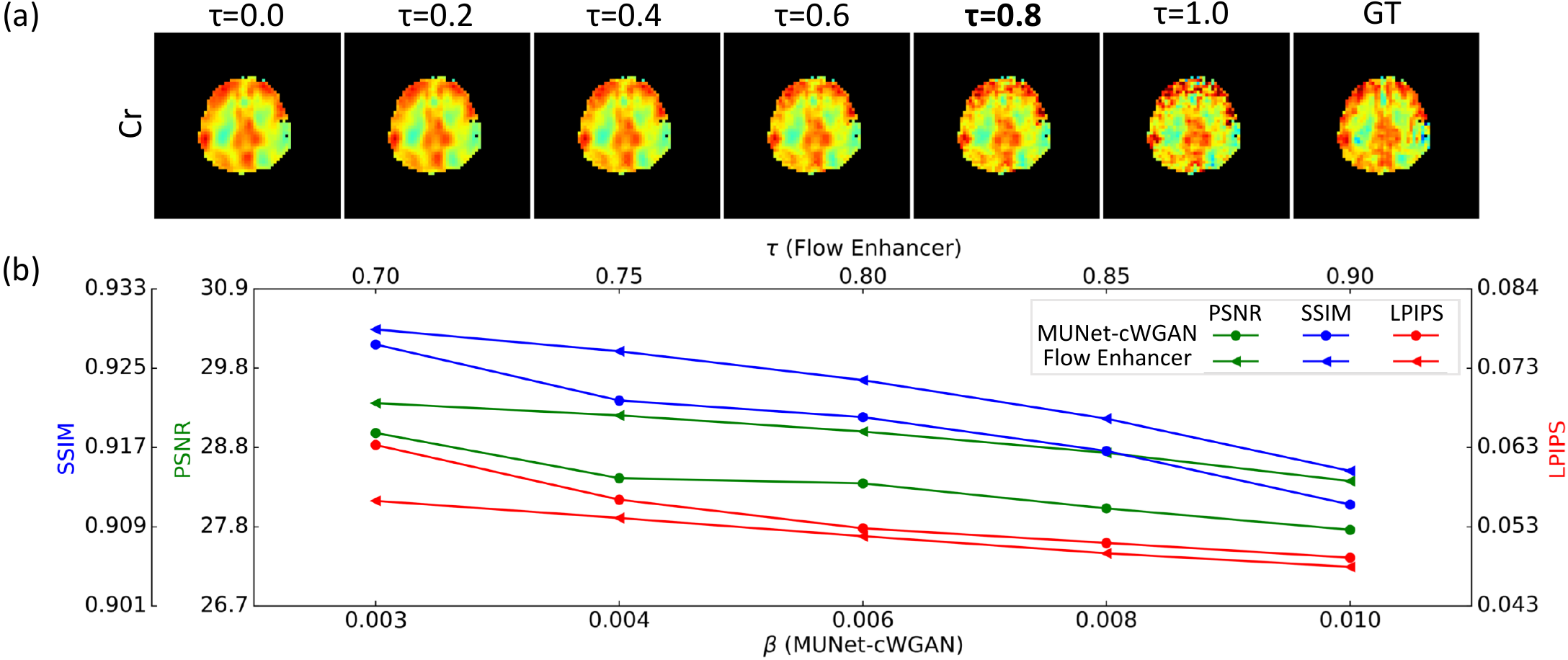}
\caption{Visual-fidelity tradeoff. (a) Sampling at different $\tau$ gives different levels of visual quality. $\tau=0.8$ (bold) gives the closest visual quality as GT. (b) PSNR($\uparrow$), SSIM($\uparrow$) and LPIPS($\downarrow$) of Flow Enhancer and MUNet-cWGAN at various levels of visual quality.
} 
\label{fig3}
\end{figure}

\begin{figure}[t]
\centering
\includegraphics[width=0.99\textwidth]{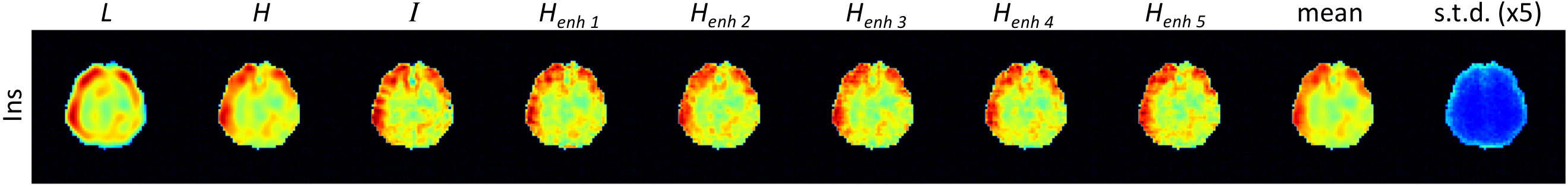}
\caption{Uncertainty estimation. From left to right: zero-filled low-resolution image $L$, blurry SR image $H$ given by the MUNet, high-resolution ground truth $I$, 5 different samples $H_{enh 1},H_{enh 2},...,H_{enh 5}$ given by the enhancer network, mean and standard deviation (s.t.d.) calculated from 100 samples. The s.t.d. map is shown in $\times5$ scale.} 
\label{fig4}
\end{figure}

\textbf{Controlling Visual-Fidelity Tradeoff via $\tau$} Our enhancer network allows tuning the tradeoff between visual quality and fidelity within the same network via $\tau$. A small value of $\tau$ gives blurry images with higher PSNR/SSIM. On the contrary, a large value of $\tau$ significantly improves the visual quality while sacrificing PSNR/SSIM. As shown in Fig. \ref{fig3}(a), $\tau=0.8$ gives the closest visual quality as the ground truth, consistent with the recommendations in previous works \cite{lugmayr2020srflow,liang2021hierarchical}. We also compare our Flow Enhancer with MUNet-cWGAN at various levels of visual quality. For MUNet-cWGAN, the tradeoff is tuned via the weight of adversarial loss, i.e. $\beta$ in $\mathcal{L}_{\text{MUNet-cWGAN}} = \mathcal{L}_{pixel} + \mathcal{L}_{structural} + \beta \mathcal{L}_{adversarial}$ \cite{dong2021high}. Note that MUNet-cWGAN requires training a separated network for each $\beta$, whereas our Flow Enhancer only needs to be trained once to obtain different levels of visual quality. As shown in Fig. \ref{fig3}(b), Flow Enhancer achieves better visual quality (LPIPS) while maintaining higher image fidelity (PSNR/SSIM) compared to MUNet-cWGAN at all levels of visual quality. 

\textbf{Uncertainty Estimation} Different from the methods based on adversarial loss, flow-based models learn a target image manifold instead of a single solution. Sampling the latent variable $\mathbf{z} \sim \mathcal{N}(\boldsymbol{\mu}(c), \tau_{0} \boldsymbol{\sigma}(c))$ generates different samples from the learned image space, of which the standard deviation can be used for uncertainty estimation of the enhanced image $H_{enh}$ \cite{denker2021conditional}. Fig. \ref{fig4} shows an example of the standard deviation map calculated from 100 samples. The uncertainty is higher around the brain periphery, which means the network observes higher variances in these regions from the training dataset. This is probably due to the lower sensitivity and stronger spectra distortion near the skull. Note that the mean image is a pixelwise average of different samples in the learned image space, therefore it looks almost identical to the blurry SR image $H$.

\section{Conclusion}
We present a flow-based enhancer network to improve the visual quality of SR MRSI. Based on the SRFlow model, we incorporated MRI prior, learnable base distribution, guide loss and DC loss to boost the performance. Results show that our method outperforms the adversarial networks and the baseline flow-based methods. Our method also allows visual quality adjustment and uncertainty estimation. The method can be extended in the future to other modalities \cite{dong2020deep}.

\subsubsection{Acknowledgements}
This work was supported by the NIH grant R01EB025840, R01CA206180 and R01NS035193. The data acquisition was supported by the Austrian Science Fund (FWF) grants KLI 646, P 30701 and P 34198.

\bibliographystyle{splncs04}
\bibliography{reference}

\end{document}